\documentclass[fleqn,12pt,twoside]{article}
\usepackage{espcrc1}

\usepackage{graphicx}
\usepackage[figuresright]{rotating}
\newcommand{\AmS}{{\protect\the\textfont2
  A\kern-.1667em\lower.5ex\hbox{M}\kern-.125emS}}

\title{Threshold $\pi^0$ photo- and electro-production in
       a meson-exchange model}

\author{S. N. Yang\address[MCSD]{Department of Physics, National Taiwan University,
        Taipei 10617, Taiwan}%
        ,
        G.-Y. Chen\addressmark,
        S. S. Kamalov\address{Laboratory of Theoretical Physics, JINR,
        Dubna 141980, Moscow Region, Russia},
        D. Drechsel\address{Institut f\"ur Kernphysik, Universit\"at
        Mainz, 55099 Mainz, Germany}, and
        L. Tiator\addressmark}

\begin{document}

\maketitle

\begin{abstract}
We show that, within a meson-exchange dynamical model describing
well most of the existing pion electromagnetic production data up
to the second resonance region, one is also able to obtain a good
agreement with the $\pi^0$ photo- and electroproduction data near
threshold. In the case of $\pi^0$ production, the effects of final
state interaction in the threshold region are nearly saturated by
single charge exchange rescattering. This indicates that in ChPT,
it might be sufficient to carry out the calculation just up to
one-loop diagrams for threshold neutral pion production.
\end{abstract}\\

Photo- and electroproduction of $\pi^0$ near threshold have been a
subject of many experimental and theoretical studies in the last decade. It was
prompted by the discrepancy between the "old" low-energy-theorem (LET)
prediction of $E_{0^+} = -2.4 \times 10^{-3}/m_\pi$ and the "new" experimental
measurements \cite{Beck90,Bergstrom}, which
yielded $E_{0^+} \sim -1.3 \times 10^{-3}/m_\pi$. The discrepancy
between LET and the experimental data was eventually resolved by the
chiral perturbation theory (ChPT) calculation
\cite{Bernard91} which showed that the loop corrections gave rise to nonanalytical
terms in  $m_\pi$. Since then precise measurements on the
$\pi^0$ electromagnetic (EM) production near threshold have been performed
\cite{Merkel02} and the ChPT  calculations to one loop $O(p^3) \,\,(O(p^4)$ in
the case of photoproduction) have been carried out in the heavy baryon
formulation \cite{Bernard96}. Nice agreement between theory and experiment was
reached not only for the $S$-wave multipoles but also for
the $P$-wave amplitudes.

 Meson-exchange models (MEM's), as in ChPT, also start from an effective
chiral Lagrangian. However, they differ from ChPT in the approach
to calculate the scattering amplitudes. In ChPT, crossing symmetry
is maintained in the perturbative field-theoretic calculation, and
the agreement between its predictions  and the data is
expected as long as the series converges. In
MEM's, the effective Lagrangian is used in the
construction of potential for use in the scattering equation. The
solutions of the scattering equation include rescattering
effects to all orders and hence unitarity is ensured, while
crossing symmetry is violated. Such models
\cite{pearce,lee91,hung,gross93,SL96,afnan99,tjon00} have been able to
provide a good description of $\pi N$ scattering lengths and phase
shifts in $S$-, $P$-, and $D$-waves up to 600 MeV pion laboratory
kinetic energy.

MEM's have been constructed for pion
EM production as well \cite{SL96,yang91}
and good agreement with the data has also been achieved up to 1.3
GeV total $\pi N$ c.m. energy. However, the predictive power of
the MEM for EM pion production near
threshold has not been fully explored even though the importance
of final state interaction (FSI) for threshold $\pi^0$
photoproduction had been demonstrated in several dynamical model
studies \cite{lee91,yang89} prior to the 1-loop
calculations of ChPT \cite{Bernard91}.

In this talk we present the predictions of the Dubna-Mainz-Taipei (DMT)
dynamical model, based on meson-exchange picture, which we recently developed in
Ref. \cite{Kamalov99} for the
threshold EM pion production and compare them with
the recent experimental data \cite{Bergstrom,Fuchs96,NIKHEF,Distler,Schmidt}
for the $S$- and $P$-wave multipoles and cross sections, and with the
results of ChPT. In our DMT model, contributions which are related to the excitation of
resonances are considered phenomenologically using standard Breit-Wigner
forms. Such an approach gives an good description of EM pion production
up to the second resonance region \cite{sabit01}.

In the dynamical model for EM pion production \cite{Yang85}, the
t-matrix is given as
\begin{eqnarray}
t_{\gamma\pi}(E)=v_{\gamma\pi}+v_{\gamma\pi}g_0(E)\,t_{\pi
N}(E)\,,\label{eq:tmatrix}
\end{eqnarray}
where $v_{\gamma\pi}$ is the $\gamma\pi$ transition potential,
$g_0$ and $t_{\pi N}$ are the $\pi N$ free propagator and
$t-$matrix, respectively, and $E$ is the total energy in the c.m.
frame. In the present study,  $t_{\pi N}$ is
obtained in a meson-exchange $\pi N$ model \cite{hung}  constructed in the
Bethe-Salpeter formalism and solved within Cooper-Jennings
reduction scheme \cite{CJ89}. Both $v_{\pi N}$ and
$v_{\gamma\pi}$  are derived from an  effective Lagrangian
containing Born terms as well as $\rho$- and $\omega$-exchange in
the $t$-channel \cite{MAID98}. For pion electroproduction
we restore gauge invariance by the substitution,
$ J_{\mu} \rightarrow J_{\mu}  - k_{\mu}(k\cdot J/k^2),$
where $J_{\mu}$ is the electromagnetic current corresponding to
the background  contribution of $v_{\gamma\pi}$.

For the physical multipoles in channel $\alpha=\{l,j,I\}$, Eq.
(\ref{eq:tmatrix}) gives~\cite{Yang85}
\begin{eqnarray}
t_{\alpha}(q_E,k)=\exp{(i\delta_{\alpha})}\,\cos{\delta_{\alpha}}
\left[ v_{\alpha}(q_E,k) + P\int_0^{~} dq'
\frac{R_{\alpha}(q_E,q')\,v_{\alpha}(q',k)}{E(q_E)-E(q')}\right]\,,
\label{eq:Tback}
\end{eqnarray}
where $\delta_{\alpha}$ and $R_{\alpha}$ are the $\pi N$ phase
shift and  reaction matrix, in channel $\alpha$, respectively,
$q_E$ is the pion on-shell momentum and $k=\mid {\bf k}\mid$ the
photon momentum. In order to ensure the convergence of the
principal value integral, we introduce a dipole-like off-shell
form factor characterizing the finite range aspect of the
potential with $\Lambda=440$ MeV.

For $\pi^0$ photoproduction, we calculate the multipole $E_{0+}$
near threshold by solving the coupled channels equation within a
basis with physical pion and nucleon masses. Results for
$Re\,E_{0+}$ are shown in Fig. 1. One sees that our results (solid
curve) agree well with the experimental data and ChPT calculations
(dash-dotted-dotted curve) \cite{Bernard96}. The FSI contributions
from  the elastic ($\pi^0 p$) and charge exchange ($\pi^+ n$)
channels, are shown by the short-dashed and dash-dotted curves,
respectively, while the dotted curve corresponds to the LET
results, i.e., without the inclusion of FSI. Our results clearly
indicate that practically all of the FSI effects originate from
the $\pi^+ n$ channel. Note that the main contribution stems from
the principal value integral of Eq. (\ref{eq:Tback}).


\begin{figure}[h]
\begin{minipage}[t]{86mm}
\centerline{\includegraphics[scale=0.3]{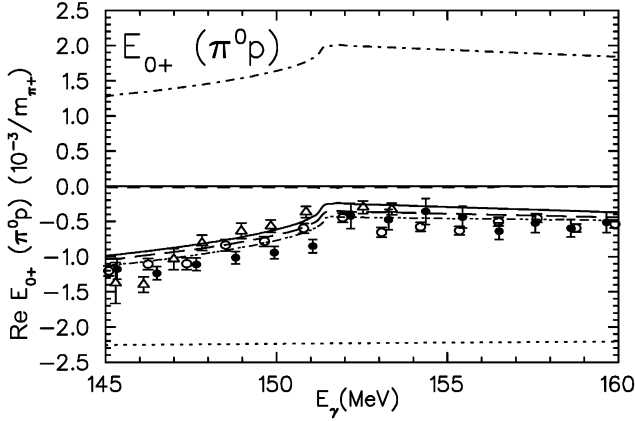}}
\end{minipage}
\hspace{\fill}
\begin{minipage}[t]{66mm}
\vspace{-6.5cm} \caption{ $ReE_{0+}$  for $\gamma p\rightarrow
\pi^0 p$. Notations are given in the text. Data points are from
($\triangle $) \protect\cite{Fuchs96}, ($\bullet$)
\protect\cite{Bergstrom}, and ($\circ$) \protect\cite{Schmidt}.}
\end{minipage}
\end{figure}


In the  approach considered above,  $t_{\pi N}$ contains the
effect of $\pi N$ rescattering to all orders. However, we have
found that only the first order rescattering contribution, i.e.
the 1-loop diagram, is important. This result is obtained by
replacing $t_{\pi N}$
 in Eq. (\ref{eq:tmatrix}) by the
$v_{\pi N}$. As can be seen in Fig. 1, the thus
obtained results given by the long-dashed curve, differ from the
full calculation by 5\% only. This indicates that the 1-loop
calculation in ChPT could be a reliable approximation for $\pi^0$
production in the threshold region.

Similar results are also obtained for the $\pi^0$ photoproduction
on neutron where 1-loop contribution with $\pi^- p$ intermediate
states is found to be large. In Table 1, the results obtained up
to tree, 1-loop, and 2-loop approximations for all four possible
pion photoproduciton channels are listed and compared to the
experiments and ChPT results. We see that for $\pi^0$ production
from both proton and neutron, it is necessary to include one-loop
contribution while tree approximation is sufficient for the
charged pion productions.


\begin{table}[htb]
Table 1. Threshold values of $E_{0+}$ ($10^{-3}/m_{\pi}$) for
different channels predicted by DMT
\newcommand{\m}{\hphantom{$-$}}
\renewcommand{\tabcolsep}{1pc} 
\renewcommand{\arraystretch}{1} 
\begin{tabular}{@{}lllllll}
\hline
        & \hspace{2mm}Tree & 1-loop & 2-loop & \hspace{4.5mm}Full & \hspace{5.5mm}ChPT & \hspace{9mm}Exp\\
\hline
$\pi^0p$&$-2.26$ &$-1.06$ &$-1.01$ &$-1.00$&$-1.1$         &$-1.33\pm0.11$\\
$\pi^+n$&\m27.72 &\m28.62 &\m28.82 &\m28.85&\m28.2\,$\pm$\,0.6 &\m28.3\,$\pm$\,0.3  \\
$\pi^0n$&\m0.46  &\m2.09  &\m2.15  &\m2.18 &\m2.13         &                \\
$\pi^-p$&$-31.65$&$-32.98$&$-33.27$&$-33.31$&$-32.7\pm0.6$&$-31.8\pm1.9$ \\
\hline
\end{tabular}
\end{table}


In  Fig. 2, we compare the predictions of our model for the
differential cross section with recent photoproduction data from
Mainz \cite{Fuchs96,Schmidt}. The dotted and solid curves are
obtained without and with FSI effects, respectively. It is seen
that both off-shell pion rescattering and cusp effects
substantially improve the agreement with the data. This indicates
that our model gives reliable predictions also for the threshold
behaviour of the $P$-waves without any additional arbitrary
parameters. A detailed comparison \cite{Kamalov01} showed that our
predictions for $P$-waves are in good overall agreement with the ChPT
predictions \cite{Bernard96} and the experimental values extracted
from recent TAPS polarization measurements \cite{Schmidt}.
However, there is a $15\% - 20\%$ difference in
$P_3=2M_{1+}+M_{1-}$ which leads to an underestimation of our
result for the photon asymmetry. Note that, in contrast to our
model, $P_3$ is essentially determined by a low energy constant in
ChPT.


\begin{figure}[htb]
\begin{minipage}[t]{76mm}
\centerline{\includegraphics[scale=0.35]{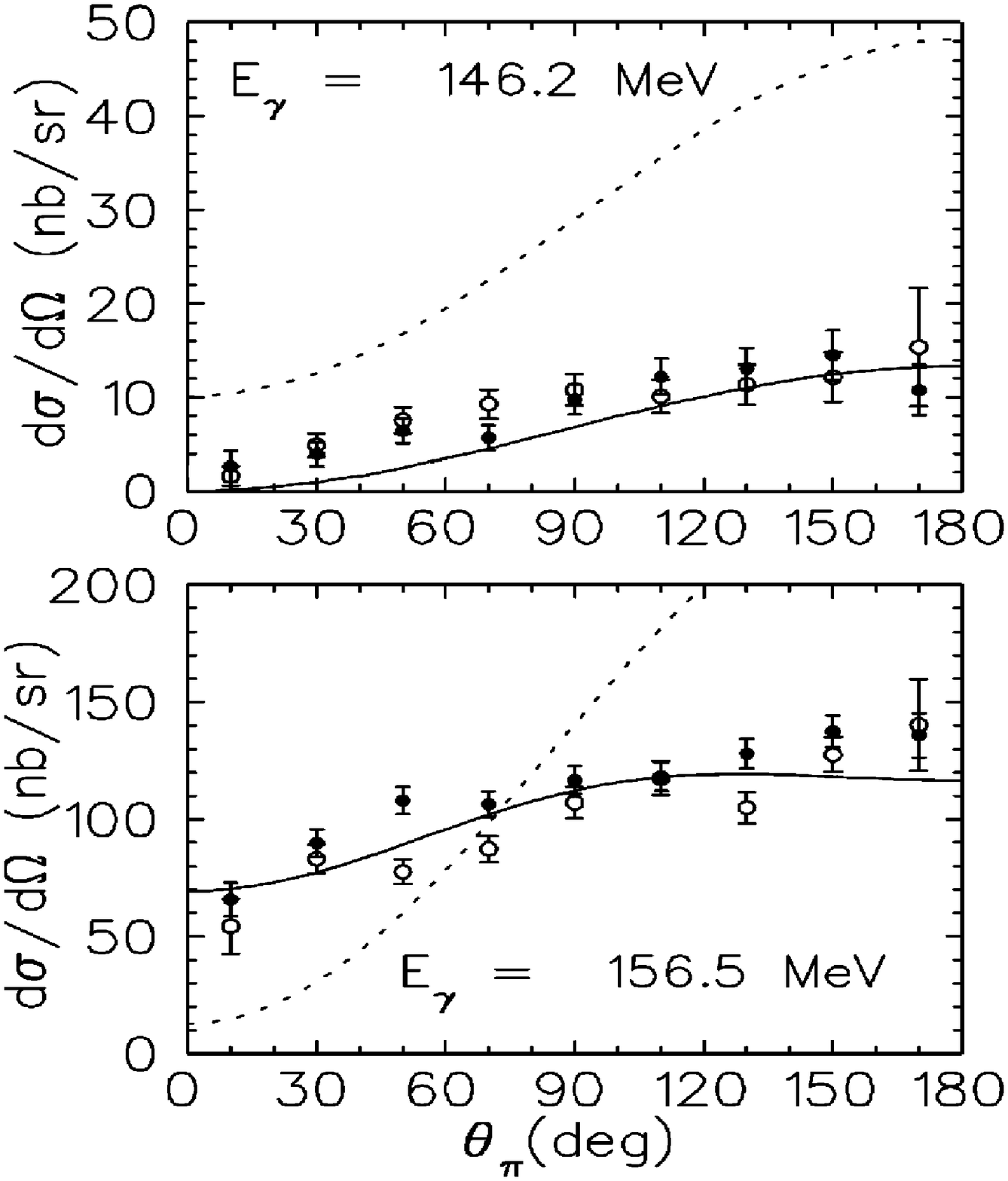}}
\caption{Differential cross sections  for $\gamma p\rightarrow
\pi^0 p$. For notations, see the text. Data points are from
($\bullet$) \protect\cite{Fuchs96} and ($\circ$)
\protect\cite{Schmidt}.}
\end{minipage}
\hspace{\fill}
\begin{minipage}[t]{76mm}
\centerline{\includegraphics[scale=0.34]{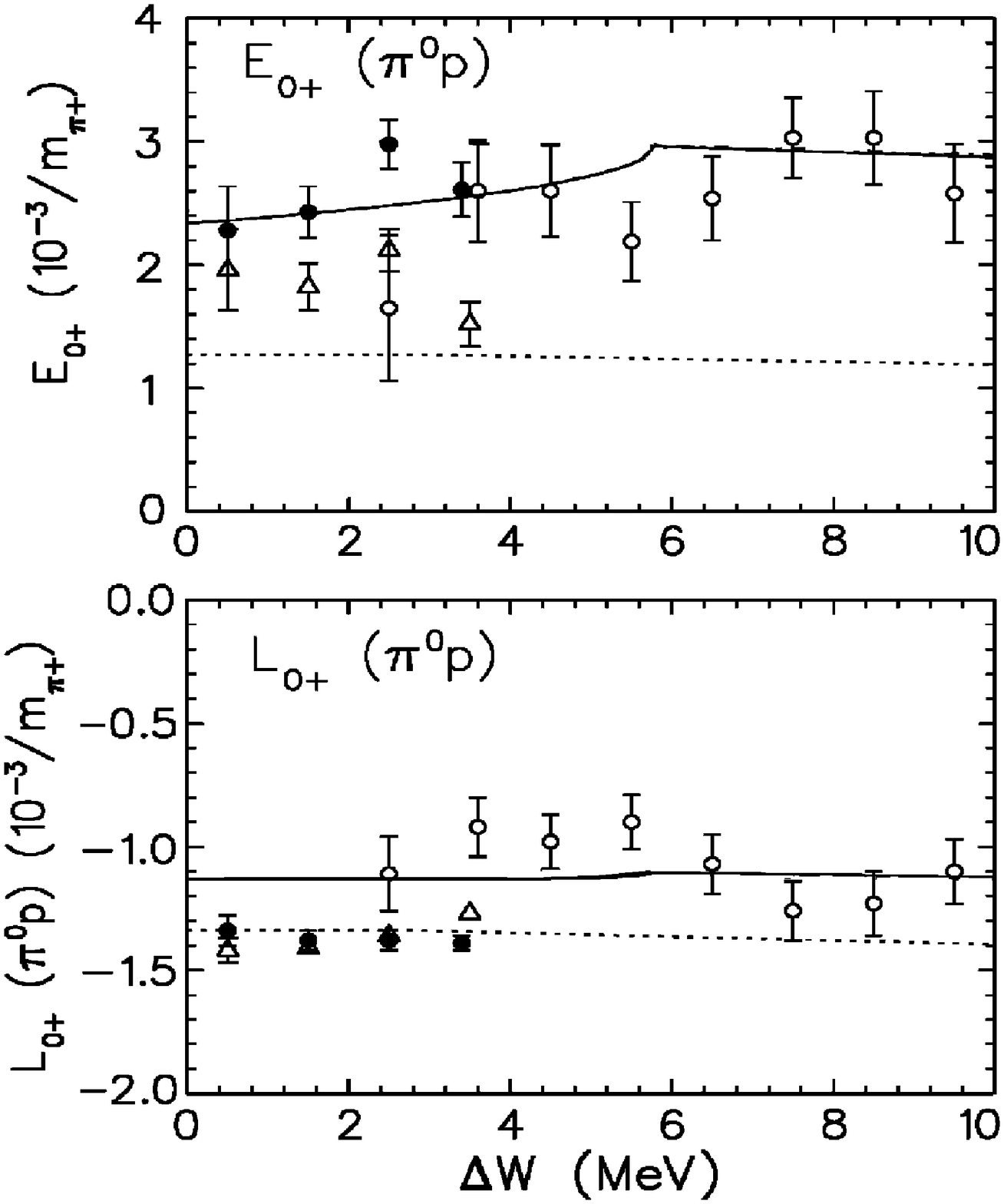}}
\caption{Re$E_{0+}$  and $L_{0+}$  at $Q^2$=0.1 (GeV/c)$^2$.
Notations same as in Fig. 3. Data points are from ($\circ$)
\protect\cite{NIKHEF} and ($\triangle$) \protect\cite{Distler}.}
\end{minipage}
\vspace{-0.5cm}
\end{figure}


Pion electroproduction provides us with information on the $Q^2 =
-k^2$ dependence of the transverse $E_{0+}$ and longitudinal
$L_{0+}$ multipoles in the threshold region. It is known that at
threshold, the $Q^2$ dependence is given mainly by the Born plus
vector meson contributions in $v_{\gamma\pi}$, as described in
Ref.~\cite{MAID98}. In Fig. 3 we show our results for the cusp and
FSI effects in the $E_{0+}$ and $L_{0+}$ multipoles for $\pi^0$
electroproduction at $Q^2=0.1$ (GeV/c)$^2$, along with the results
of the multipole analysis from NIKHEF \cite{NIKHEF} and Mainz
\cite{Distler}. Note that results of both groups were obtained
using the $P$-wave predictions given by ChPT. However, there exist
substantial differences between the $P$-wave predictions of ChPT
and DMT model at finite $Q^2$. To understand the consequence of
these differences,  we have made a new analysis of the Mainz data
\cite{Distler} for the differential cross sections, using DMT
prediction for the $P$-wave multipoles instead. The $S$-wave
multipoles extracted this way  are also shown in Fig. 3 by solid
circles.  We see that the results of such a new analysis give a
$E_{0+}$ multipole closer to the NIKHEF data and in better
agreement with our dynamical model prediction. However, the
results of our new analysis for the longitudinal $L_{0+}$
multipole stay practically unchanged from the values found in the
previous analyses. Note that the dynamical model prediction for
$L_{0+}$  again agrees much better with the NIKHEF data.


\begin{figure}[h]
\centerline{\includegraphics[scale=0.5]{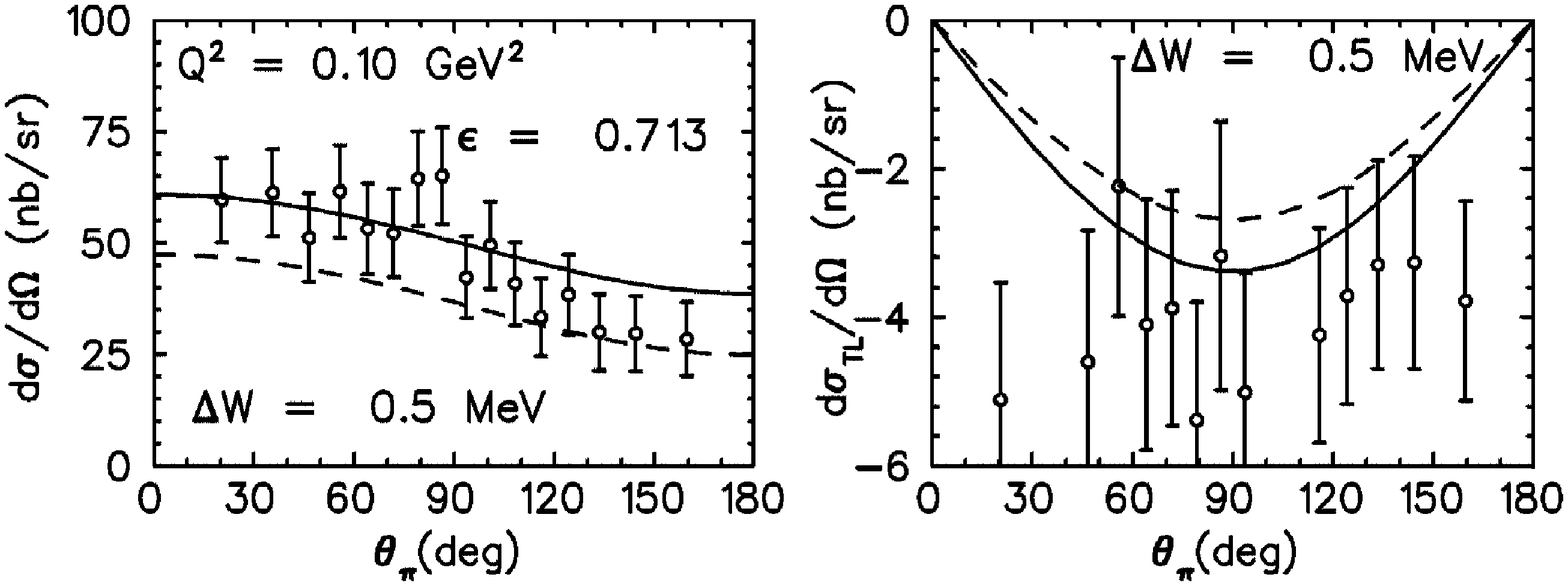}}
\caption{$d\sigma/d\Omega$  and $d\sigma_{TL}/d\Omega$  at
$Q^2$=0.1 (GeV/c)$^2$ and $\epsilon=0.713$, at $\Delta
W=W-W_{thr}^{\pi^0p}=0.5 MeV$. For notations, see the text. Data
points are from Ref. \protect\cite{Distler}.}
\end{figure}


In Fig. 4, DMT model predictions (dashed curves) are compared with
the Mainz experimental data \cite{Distler} for the unpolarized
cross sections $d\sigma/d\Omega$, and for the
longitudinal-transverse cross section $d\sigma_{TL}/d\Omega$.
Overall, the agreement is good. The solid curves are the results
of our best fit at fixed energies (local fit) obtained by varying
only the $E_{0+}$ and $L_{0+}$ multipoles. We have found that the
differences between the solid and dashed curves in Fig. 4 are
mostly due to the difference in the $L_{0+}$ multipole (see also
Fig. 3).

In summary, we have shown that within a meson-exchange dynamical
model \cite{Kamalov99}, one is able to describe pion photo- and
electroproduction in the threshold region in good agreement with
the data. The model has been demonstrated to give a good
description of most of the existing pion electromagnetic
production data up to the second resonance region \cite{sabit01}.
The success of such a model at intermediate energies is perhaps
not surprising since unitarity plays an important role there.
However, it is not {\it a priori} clear that our model should also
work well near threshold, even though we do start from an
effective chiral Lagrangian. In principle, crossing symmetry is
violated and the well-defined power counting scheme in ChPT is
lost by rescatterings.  On the other hand, MEM's
\cite{afnan99,tjon00} have also been shown to give a good
description of low energy $\pi N$ data, in addition to an
excellent agreement with the data at higher energies. It is
therefore assuring that similar success can also be achieved for
the pion EM production.

Finally, we found that the effects of FSI in the threshold region
and in the case of $\pi^0$ production, are nearly saturated by the
single rescattering term. Therefore, the existing one-loop
calculations in ChPT  could be a good approximation to threshold
$\pi^0$ production.

\end{document}